\documentclass[sigconf,balance=false]{acmart}

\AtBeginDocument{%
  \providecommand\BibTeX{{%
    \normalfont B\kern-0.5em{\scshape i\kern-0.25em b}\kern-0.8em\TeX}}}

\setcopyright{acmcopyright}
\copyrightyear{2023}
\acmYear{2023}
\acmDOI{XXXXXXX.XXXXXXX}

\acmConference[CONCATENATE - HRI]{}{March 13, 2023}{Stockholm, SE}
\acmBooktitle{Proceedings of 18th ACM/IEEE International Conference on Human-Robot Interaction (HRI 2023),
 March 13--16, 2023, Stockolm, SE} 
\acmPrice{15.00}
\acmISBN{978-1-4503-XXXX-X/18/06}

\begin{document}

\title{The mentor-child paradigm for individuals with autism spectrum disorders}

\author{Marion Dubois-Sage}
\orcid{0000-0003-0928-2350}
\affiliation{%
  \institution{Laboratoire CHArt-UP8 RNSR 200515259U}
  \streetaddress{2 rue de la liberté}
  \city{Saint-Denis}
  \country{France}}
\affiliation{%
  \institution{Université Paris 8}
  \streetaddress{2 rue de la liberté}
  \city{Saint-Denis}
  \country{France}
}

\author{Baptiste Jacquet}
\orcid{0000-0001-5310-2789}
\affiliation{%
  \institution{Laboratoire CHArt-UP8 RNSR 200515259U}
  \streetaddress{2 rue de la liberté}
  \city{Saint-Denis}
  \country{France}}
\affiliation{%
  \institution{Université Paris 8}
  \streetaddress{2 rue de la liberté}
  \city{Saint-Denis}
  \country{France}}

\author{Frank Jamet}
\orcid{0000-0002-6630-7633}
\affiliation{%
  \institution{Laboratoire CHArt-UP8 RNSR 200515259U}
  \streetaddress{2 rue de la liberté}
  \city{Saint-Denis}
  \country{France}}
\affiliation{%
  \institution{CY Cergy Université}
  \city{Cergy-Pontoise}
  \country{France}
}

\author{Jean Baratgin}
\authornotemark[1]
\orcid{0000-0001-9566-486X}
\affiliation{%
  \institution{Laboratoire CHArt-UP8 RNSR 200515259U}
  \streetaddress{2 rue de la liberté}
  \city{Saint-Denis}
  \country{France}}
\affiliation{%
 \institution{Université Paris 8}
 \streetaddress{2 rue de la liberté}
 \city{Saint-Denis}
 \country{France}}
 \email{jean.baratgin@univ-paris8.fr}





\renewcommand{\shortauthors}{Dubois-Sage, et al.}


\begin{abstract}
    Our aim is to analyze the relevance of the mentor-child paradigm with a robot for individuals with Autism Spectrum Disorders, and the adaptations required. This method could allow a more reliable evaluation of the socio-cognitive abilities of individuals with autism, which may have been underestimated due to pragmatic factors.
\end{abstract}




\received{20 January 2023}
\received[revised]{1st March 2023}
\received[accepted]{1st March 2023}

\maketitle

\section{Introduction}

The present article aims at highlighting the interest of the mentor-child paradigm \citep{Jamet18b} for individuals with Autism Spectrum Disorders (ASD). This experimental paradigm, which consists in replacing the interaction of a child with an adult by an interaction with a robot, has already shown its effectiveness in re-evaluating the performance of neurotypical children previously underestimated \citep{baratgin2020}. It could also have benefits for children with ASD, both experimentally by allowing a more detailed study of their socio-cognitive abilities, and developmentally by providing a supportive environment for learning.

Autism is a neurodevelopmental disorder characterized by restricted or repetitive patterns of behavior, interests, or activities, as well as deficits in communication and social interactions, \citep{american_psychiatric_association2013}. Therefore, although the mentor-child paradigm appears to be effective with neurotypical children, it may need to be adapted for implementation with individuals with autism. 
The way they perceive and interact with robots might differ from that of neurotypical individuals.
 Since the processing of robot features - both appearance and behavior - depends on the cognitive abilities of the user \citep{johnson2003}, and since individuals with ASD may have cognitive impairments, they will not necessarily interpret cues in the same way \citep{schadenberg2020}.

\section{AMBIGUITY IN CONVERSATIONAL TESTS}


Many psychological tests aimed at assessing children's socio-cognitive development are based on conversational interaction, i.e., they involve a dialogue between an experimenter and a child. Four common features of these tests have been identified by \citet{baratgin2021} and \cite{baratgin21a}: 1) A problem situation is introduced to the child by the presence of a physical device (e.g., pictures or objects); 2) The adult examiner (or experimenter) initiates a conversation with the child to explain the problem situation (e.g., tells the child a story based on pictures); 3) The experimenter asks the child a test question. The child's response to this target question will determine whether or not the child has acquired the concept being studied;\footnote{This response can be verbal - with an oral answer - or non-verbal - with a gesture.} 4) These studies take place mostly in a room provided at the child's school which places them in a school context. In other words, when the child has to answer a question asked by the experimenter, they do not differentiate it from a question from their teacher. This teaching context determines the role assigned to each agent (the child has the role of student, and the adult has the role of the teacher).


In these conversational tests, performance can be impacted by the pragmatics of language \citep{politzer2016}. Indeed, success in this type of task depends on the child's response to a test question. When asked a question, the child will seek to infer the communicative intent of the experimenter in order to best answer it. However, this question may be ambiguous, i.e., it may correspond to several possible interpretations. In such a situation, according to relevance theory \citep{sperber2001}, we select the interpretation that seems most relevant to the context in which the question is stated. To provide the correct answer to the test question, the child must have the same interpretation of the question as the experimenter. If their interpretations differ, the child will not provide the expected answer, but this does not mean that the child does not know the correct answer. Thus, the child's interpretation of the test question is constrained by their developmental level: children will attribute different intentions to the experimenter depending on their age. The poor performance of typically developing young children on this type of task would therefore be due to a discrepancy in interpretation between the experimenter and the child about how to interpret the test question \citep{politzer2016}. It has also been shown that performance in various interactions with digital tools is influenced by pragmatic factors.\citep[for example see][]{masson2015,masson2016,Masson17a,jacquet2018,jacquet2019}.


\section{The mentor-child paradigm with typically developing individuals}
The mentor-child paradigm does, however, reduce the pragmatic ambiguity of conversational tests. This method consists of replacing the conversation with the adult experimenter in the original test with a conversation with a robot, and inverting the social statuses: the child assumes the role of teacher towards a NAO robot presented as an ignorant student whom they must educate. The ignorance of the NAO robot is reinforced during the conversation, as it tells the child that it does not know anything and that it needs their help to learn.  In this new experimental context, the examiner who asks the test question is not a knowing adult, but a naive robot. Conversely, when the child answers the question, they take on the role of a knowing teacher rather than an ignorant student. When the test question is asked by an ignorant entity, such as the NAO robot, the new context in which the question is stated makes it easier to select the expected interpretation.

The tendency of neurotypical children to treat robots as social agents \citep{di_dio2020}, and to help them, \citep{martin2020a} facilitates the establishment of this teacher/student relationship between child and robot. Nevertheless, individuals with ASD might present different behaviors when interacting with robots, which would imply adaptations of the paradigm (a point that will be discussed later, cf. section 4).


In typically developing children, the presence of a robot as an experimenter thus allows us to observe a better performance of 3-4-year-olds in a conversational test such as the false belief task \citep{baratgin2020}. Children under 4 years have the ability to attribute false beliefs to others but are not able to infer the expected interpretation of the test question due to insufficient pragmatic abilities. On the contrary, for older children and adults, who have more developed pragmatic abilities through social experience, it would be less costly to understand what the test question refers to in this context, and to identify the experimenter's actual expectation, allowing them to give the correct answer. These results confirm that the pragmatic factors identified by \citet{politzer2016} are a barrier to young children's success in the original false belief test. The pragmatic factors involved in the conversational tests would therefore encourage underestimation of neurotypical children's social-cognitive abilities, and could also impede success in other populations.


Therefore, it seems relevant to re-examine the performance of children with atypical development, particularly that of individuals with ASD, on conversational tests. Indeed, individuals with ASD may have deficits in language pragmatics \citep{angeleri2016}. It is possible that the limited pragmatic abilities of individuals with ASD may impact their performance on conversational tests in the same way that it impacts the performance of neurotypical children. The ambiguity of the test question and the context of enunciation, added to possible difficulties in language pragmatics, could hinder their success, and lead to an underestimation of their socio-cognitive abilities. Thus, the mentor-child paradigm is of major interest to re-evaluate the abilities of individuals with autism in different socio-cognitive domains, such as social cognition or mathematical logic.


\section{Adaptation of the mentor-child paradigm for individuals with autism}


The mentor-child paradigm assumes that the child is able to take on the role of a teacher educating a naive robot. More precisely, we suppose that it implies several prerequisites: 1) that the child sees the robot as a social agent; 2) that the child is willing to help the robot; 3) that the child understands the concept of ignorance, and what is expected of a student; 4) that the child understands what is expected of a teacher, i.e., to transmit knowledge. These different points will be examined in order to determine the necessary adaptations in this paradigm for children and adults with ASD.


\subsection{Interaction with robots as social agents and helping behavior}
To establish a master-student relationship between neurotypical children and the robot, we rely on their tendency to treat the robot as a social agent driven by desires \citep{chernyak2016} and intentions \citep{martin2020b}, and their willingness to help the robot \citep{martin2020a}. However, we can question whether children and adults with autism have a comparable conception of robots as neurotypical individuals and whether they exhibit similar interaction and helping behaviors. Indeed, motivation to help others is related to social skills \citep{komeda2019}, impacted in ASD.


Autistic and neurotypical children aged 5-7 years categorize robots as toys \citep{peca2014, zhang2019a}. Different studies which have introduced the NAO robot to typically developing children \citep{nijssen2021, van_straten2022}, but also to children with ASD \citep{zhang2019b, marino2020,rakhymbayeva2021} show that this robot is well accepted by children, regardless of their developmental type. The choice of the NAO robot, the most widely used for interaction with children with ASD \citep{saleh2021,raptopoulou2021}, thus seems appropriate for the mentor-child paradigm. Like neurotypical children \citep{di_dio2020}, special needs children (including children with ASD) seem to interact with a robot in the same way as they would with a human \citep{wood2013}. They may even prefer interaction with a robot over a human \citep{wood2013, simlesa2022}. In a robot-facing interaction, children with autism show prosocial behaviors similar to those of neurotypical children, such as eye contact, vocalizations, and smiles toward the robot \citep{scassellati2007}. In the natural context of interaction (i.e., with a human), such behaviors are rarely observed in children with ASD.


We have seen that neurotypical children can spontaneously help a robot \citep{martin2020a}, but it is possible that this behavior differs in individuals with ASD. Indeed, adolescents with autism aged 10-13 years would show less helping behavior compared to adolescents with Down syndrome \citep{sigman1999} and neurotypical adolescents \citep{oconnor2019}. Yet, children with autism aged 2-5 years show a similar pattern of helping behaviors as children with developmental delay \citep{liebal2008} or even more helping behaviors compared to neurotypical children at age 3-6 years \citep{paulus2017}. Another study of autistic and neurotypical children of the same age indicates similar performance on a spontaneous helping task \citep{dunfield2019}. Thus, individuals with autism are able to express appropriate prosocial behaviors despite their social difficulties and are able to identify a situation in which someone needs help \citep{dunfield2019}. In the mentor-child paradigm, then, children with autism might strive to help the robot in the same way that neurotypical children do, although to our knowledge there are no studies examining the helping behaviors of children with autism toward a robot.


In \citep{komeda2019}, several scenarios were presented to neurotypical adults and adults with ASD featuring an agent with either autistic or neurotypical characteristics. Results showed that neurotypical adults were more motivated to help the neurotypical agent compared to adults with ASD. Conversely, for the agent with autistic characteristics, both groups show similar motivation to help. After controlling for some variables (Alexithymia \footnote{Alexithymia is the inability to recognize and describe emotions verbally \citep{taylor1984}.} Score and Autism Quotient) in the group of adults with ASD, the results show a higher motivation to help the agent with autistic characteristics. Thus, presenting a robot as having autistic characteristics (e.g., comprehension difficulties in social interactions) in the mentor-child paradigm could increase the helping behaviors of children with ASD towards it.

\subsection{Teacher status with a naive student}
To emphasize NAO's ignorance and thus reinforce his naive student status and the child's teacher posture, the robot tells them that it knows nothing and needs their help to learn. However, we can ask how individuals with ASD perceive ignorance in others, and whether they are likely to transmit knowledge.


It appears that children and adolescents with ASD (ages 7-18) have a delay in understanding the concept of ignorance in others \citep{leslie1988,perner1989}, an ability that is acquired around age 3-4 in neurotypical children \citep{hogrefe1986}. Still, individuals with ASD are able to transmit knowledge to a robot in an interactive learning paradigm. Authors have proposed to neurotypical children and children with ASD to teach animal names to a robot (KASPAR) by providing feedback to each of its propositions \citep{zaraki2020}. The results show similar robot learning in both groups, and this interactive robot learning paradigm appears to be as effective and accessible for children with autism as it is for neurotypical children. In another study on collaborative learning, children with ASD took on the role of teacher to a robot that made reading errors. They were able to correct the robot's mistakes and explain the correct answer \citep{Jimenez2017}. Thus, individuals with autism appear to be able to help an ignorant robot acquire knowledge, suggesting that they may adopt the status of a teacher to a student, as implied by the mentor-child paradigm.


To summarize, the mentor-child paradigm with a NAO robot appears to be adaptable to children with ASD as well as neurotypical children: they interact with robots as they do when interacting with humans \citep{wood2013} or even prefer these interactions \citep{wood2013, simlesa2022}; they show spontaneous helping behavior comparable to that of neurotypical children \citep{dunfield2019}; and they can impart knowledge to robots \citep{zaraki2020}. This last point assumes that they understand—at least partially – the concept of ignorance, and that they are able to adopt a teaching role with a student robot. The NAO robot, among the most used with this population \citep{saleh2021,raptopoulou2021}, seems to be a relevant choice for interaction with individuals with ASD.


The applicability of this paradigm to individuals with ASD may provide an experimental setting that is neutralized from factors of language pragmatics, which would allow for a re-examination of their performance on classical psychological tests. The classical conversational context may have led to an underestimation of their abilities. However, one point needs to be emphasized: while this paradigm has been shown to be effective and appropriate with neurotypical children, its benefits to children and adults with ASD remain to be proven through experimental studies.

Moreover, children with ASD show a variety of responses when interacting with a robot, which could be explained, at least in part, by the degree of severity of their disorder \citep{telisheva2022}. Children with ASD with the best language skills and social functioning initiate more functional interactions (imitation, dialogue) with the robot \citep{schadenberg2020}. It is important to note the significant differences that can be observed in the symptomatology of individuals with ASD, particularly in cognitive functioning - which can range from profound impairment to high intellectual ability \citep{grzadzinski2013} - and language functioning - with some individuals with ASD being non-verbal, and others having preserved language abilities (despite difficulties in the pragmatic component of language) \citep{schadenberg2020}. This highlights the need to adapt the type of activity as much as possible to the characteristics of each child, in order to provide personalized interventions. However, despite the proposed adaptations and consideration of individual characteristics, it is possible that the mentor-child paradigm may not be suitable for some individuals with ASD.


\section{A paradigm to support the socio-cognitive development of ASD individuals} 
In addition to its usefulness in psychological research to re-examine the performance of individuals with ASD on social-cognitive tests, this method could also be used to support the socio-cognitive development of these children.


The mentor-child paradigm can be used as a reinforcer \citep{zaraki2020}, but the robot can also act as an emulator, to promote vicarious learning, i.e., through the imitation of a peer performing the behavior to be learned \citep{saadatzi2018}. This study with children with autism aged 6-8 years involved learning words through a virtual teacher alongside a humanoid robot (NAO) presented as the child's peer. The teacher could offer the same word to the child and the robot, or submit two different words. The participants made fewer errors in learning a word when the robot had to learn the same word. We also observe that through a process of vicarious learning, the children integrated 94\% of the words that were proposed only to the robot. The presence of a robot thus seems to increase the motivation of children with autism to acquire knowledge. Therefore, the mentor-child paradigm could also accompany the learning of children with autism.


\section{Conclusion}
The mentor-child paradigm with a NAO robot could be a particularly relevant experimental method with individuals with ASD, as it could lead to greater success in conversational tests, as is the case with young neurotypical children \citep{baratgin2020}. For the method to be suitable for individuals with ASD, few modifications will be necessary, as long as the difficulty of the interaction content is appropriate for \citep{scassellati2018,clabaugh2019} the child. However, it may be appropriate to present the robot as having social difficulties, as this would increase motivation to help in children with autism \citep{komeda2019}. Children with ASD show similar helping abilities to neurotypical children \citep{dunfield2019}, and appear to interact with robots as they do with humans \citep{wood2013}. They can also impart knowledge to a robot \citep{zaraki2020}, suggesting that they would be able to take on the role of a robot's teacher in the same way as neurotypical children, despite their lower understanding of the concept of ignorance.


Furthermore, considering the assumed preference for interacting with a robot over a human \citep{wood2013, simlesa2022}, the mentor-child paradigm could be used from a developmental perspective, especially to support school activities, since children with ASD can learn through imitation of a robot \citep{saadatzi2018}.


Thus, the mentor-child paradigm would not only reveal better socio-cognitive skills in individuals with ASD, especially during conversational tests but could also promote the development of these skills. This new experimental method is therefore of great interest to the autistic population. However, experimental studies are needed to demonstrate whether this paradigm is suitable for individuals with ASD and whether it actually provides benefits.

\begin{acks}
We would like to thank the P-A-R-I-S association for funding this paper.
\end{acks}

\bibliographystyle{ACM-Reference-Format}
\bibliography{sample-base}

\end{document}